\documentclass[%
 reprint,
 superscriptaddress,
 amsmath,amssymb,
 aps,
 pra,
]{revtex4-1}

\usepackage{graphicx}
\usepackage{dcolumn}
\usepackage{bm}
\usepackage{hyperref}

\usepackage{color}
\definecolor{rltred}{rgb}{0.75,0,0}
\definecolor{rltgreen}{rgb}{0,0.6,0}
\definecolor{rltblue}{rgb}{0.3,0.3,1}

\usepackage{placeins}

\hypersetup{colorlinks,%
    hypertexnames=true,%
    linkcolor=rltred,%
    citecolor=rltgreen,%
    linktocpage=true}

\newcommand{\XUV}{{\scriptscriptstyle\mathrm{XUV}}}
\newcommand{\IR}{{\scriptscriptstyle\mathrm{IR}}}

\newcommand{\EWS}{{\scriptscriptstyle\mathrm{EWS}}}

\newcommand{\cc}{{\scriptscriptstyle\mathrm{CC}}}

\begin{document}


\title{Angle-resolved time delays for shake-up ionization of helium}

\author{Stefan Donsa}
\email{stefan.donsa@tuwien.ac.at}
\affiliation{Institute for Theoretical Physics, Vienna University of Technology,
    Wiedner Hauptstra\ss e 8-10/136, 1040 Vienna, Austria, EU}
\author{Manuel Ederer}
\affiliation{Institute for Theoretical Physics, Vienna University of Technology,
    Wiedner Hauptstra\ss e 8-10/136, 1040 Vienna, Austria, EU}
\author{Renate Pazourek}
\affiliation{Institute for Theoretical Physics, Vienna University of Technology,
    Wiedner Hauptstra\ss e 8-10/136, 1040 Vienna, Austria, EU}
\author{Joachim Burgd\"orfer}
\affiliation{Institute for Theoretical Physics, Vienna University of Technology,
    Wiedner Hauptstra\ss e 8-10/136, 1040 Vienna, Austria, EU}
\author{Iva B\v{r}ezinov\'a}
\affiliation{Institute for Theoretical Physics, Vienna University of Technology,
    Wiedner Hauptstra\ss e 8-10/136, 1040 Vienna, Austria, EU}

\date{\today}
\begin{abstract}
Recent angle-resolved RABBITT experiments have shown that the photoionization time delay depends on the emission angle of the photoelectron.
In this work we demonstrate that for photoemission from helium accompanied by shake-up (correlation satellites), the angular variation of the time delay is dramatically enhanced by the dipolar coupling between the photoelectron and the highly polarizable bound electron in the IR field.
We show that the additivity rule for the time delays due to the atomic potential, the continuum-continuum (cc) coupling by the IR field, and due to this two-electron process remains valid for angle-resolved RABBITT. 
Our results are expected to be also applicable to other multi-electron systems that are highly polarizable or feature a permanent dipole moment.
\end{abstract}

\maketitle
\section{Introduction}
Progress in the development of coherent light sources has fostered the development of new protocols to monitor and steer electronic and molecular motion \cite{BraKra2000}.
High harmonic generation \cite{McpGibJar1987,FerLhuLi1988,LewBalIva1994} has  enabled the generation of isolated \cite{HenKieSpi2001} and trains of light pulses \cite{GloSchChi1996,SchBreAgo1994,PauTomBre2001} with duration in the attosecond time domain.
These developments enabled schemes to control valence electrons on their natural time scale (attoseconds) \cite{KraIva2009} and to directly measure the phase imprinted on the electronic wavepacket generated by the photoelectric effect.
Theoretical insights have revealed that the phase of the ejected electrons can be directly related to the  Eisenbud-Wigner-Smith (EWS) scattering time delay $\tau_{\EWS}$ \cite{PazNagBur2015}. 
The phase of the wavepacket and, thus, the photoionization time delay strongly depends on the underlying physical process.
Scenarios investigated so far include electron emission originating from atoms \cite{SchFieKar2010,KluDahGis2011,PazFeiNag2012,GueKroBal2014,IsiSquBus2017,JimMarArg2016,StoCavDon2018,DonDouBur2019}, molecules \cite{HupJorBay2016,VosCatPat2018,BisShuFor2020}, and solids \cite{CavMulUph2007},  tunneling ionization \cite{EckPfeCir2008}, above-threshold ionization \cite{ZipNatBuc2014}, and double ionization \cite{PazNagBur2015a}.
\\
First experiments and calculations investigating photoionization time delays have either  been performed by the attosecond streak camera \cite{CavMulUph2007,SchFieKar2010,NagPazFei2011} which records photoelectrons along the polarization axis of the IR streaking field, or by RABBITT (reconstruction of attosecond bursts by interference of two-photon transtions) \cite{KluDahGis2011,GueKroBal2014,PalDahKhe2014,SabHeuBog2015}, measuring angle-integrated photoelectron spectra.
Alternatively, angular streaking, employing circularly polarized pulses has been used to extract ionization phases and timing information from the electron angular distribution, e.g.~\cite{EckPfeCir2008,SaiXuWan2019,HeRuiHe2016,ArmClaBen2020}.
More recently, measurements of time delays by RABBITT as a function of the emission angle of the electrons with respect to the laser polarization direction have become available \cite{HeuJimCir2016,CirMarHeu2018,BusVinZho2018,FucDouDon2020}, revealing a pronounced angle dependence for large emission angles.
Calculations have been able to reproduce this dependence for different noble gas atoms \cite{Iva2011,DahLin2014,WaeMosPav2014,IvaKhe2017,BraNasKhe2018,Hoc2017}.
The variation of the time delay for large emission angles cannot be attributed to the EWS delay $\tau_{\EWS}$  associated with the XUV induced bound-free transition \cite{IvaKhe2017,BraNasKhe2018} but rather to the additional time delay  associated with the absorption (emission) of the IR photon in a Coulomb field, the continuum-continuum delay $\tau_{\cc}$, which is inherent to the RABBITT protocol \cite{KluDahGis2011}.
Recent results show that the angle dependence of $\tau_{\cc}$ can be viewed as a result of the IR induced partial-wave interference and the phase shift of the outgoing electron in the Coulomb field \cite{DahLhuMaq2012,FucDouDon2020,Fan1985,BusVinZho2018}. 
\\
For several atomic species the time delay observed in photoionization was found to be the sum of EWS delay $\tau_{\EWS}$ and continuum-continuum delay $\tau_{\cc}$ (or, the Coulomb-laser coupling (CLC) delay for streaking) \cite{PazNagBur2015,NagPazFei2011,KluDahGis2011}. \\
For photoionization accompanied by shake-up of the second electron, the socalled correlation satellites, theoretical predictions \cite{PazFeiNag2012} and experiments \cite{OssSieShi2017} have shown that the coupling of the transient dipole moment of the bound electron to the IR field imprints an additional phase shift on the correlated two-electron wave function which manifests itself as an additional correlation contribution $\tau_{e-e}$ to the time delay of the emitted electron. 
So far, detection of this observable was limited to the emission direction along the polarization axis in the streaking geometry \cite{OssSieShi2017}. 
RABBITT offers the opportunity to study the angular dependence of the time delay of these correlation satellites. 
Moreover, the superior energy resolution of RABBITT allows to spectrally resolve \cite{IsiSquBus2017} individual shake-up states He$^+(n)$ with $n=2,3$.
While the correlation delay $\tau_{e-e}$ for atoms is mostly due to transient polarization of the target, molecules featuring a permanent dipole moment are expected to give rise to an even stronger angle dependence of the time delay \cite{BagMad2010,BisShuFor2020}.
\\
In this work we analyze the angular dependence of time delays in the presence of shake-up excitation. 
We show that the additivity of three distinct contributions to the time delay, the Eisenbud-Wigner-Smith time delay $\tau_{\EWS}$ for bound-free transitions in the absence of the probing IR field, the continuum-continuum contribution $\tau_{\cc}$ due to transitions induced by the IR field, and the two-electron correlation delay $\tau_{e-e}$ remains valid for angle-resolved RABBITT. 
Disentangling the different contributions, we find that $\tau_{e-e}$ provides by far the dominant contribution to the angle dependence for moderate emission angles $\theta$ relative to the polarization axis $(\theta \leq 70^{\circ})$ where the emission probability is still large.
Furthermore, we compare time delays extracted from angle-integrated RABBITT traces to time delays for angle-resolved RABBITT traces.
We show that RABBITT spectra restricted to forward direction ($\theta=0^{\circ}$) yield the same delay as attosecond streaking.
This demonstrates that attosecond streaking and RABBITT allow access to the same physical quantity even for complex multi-electron systems when electron-electron correlations are strong.\\
The paper is organized as follows: In Sec.~\ref{sec:rabbitt_preliminaries} we briefly review the RABBITT protocol for observing angle-resolved time delays.
In  Sec.~\ref{sec:angle_dependence_tau_r}, we compare and discuss the angle dependence of the EWS delay $\tau_{\EWS}$, the continuum-continuum delay $\tau_{\cc}$, and of the dipole-induced correlation delay $\tau_{e-e}$ in helium.
Finally, a comparison between angle-integrated and angle-resolved delays for shake-up ionization measured by RABBITT and those measured by recent attosecond streaking experiments will be presented in Sec.~\ref{sec:integrated_vs_resolved}, followed by concluding remarks in Sec.~\ref{sec:conclusion}.
Numerical details of the simulations can be found in the appendix.

\section{Angle-resolved photoionization time delays observed by RABBITT}  \label{sec:rabbitt_preliminaries}

Up to now, the dipole-induced correlation delay $\tau_{e-e}$ in photoionization accompanied by shake-up in helium was calculated and observed within an attosecond streaking protocol \cite{OssSieShi2017}. 
Therefore, the accessible information was restricted to the emission in forward direction $\left( \theta=0^{\circ} \right)$ relative to the polarization axis of the streaking IR field.
The RABBITT protocol provides an attractive alternative as it offers two advantages: higher spectral selectivity thereby enabling the resolution of nearby shake-up lines \cite{IsiSquBus2017} as well as both angle-integrated and angle-resolved time delay measurements \cite{KluDahGis2011,GueKroBal2014,HeuJimCir2016}.
We therefore briefly review the underlying concepts of angle-resolved time delays as detected by RABBITT for the well-known EWS and cc delays before analyzing in detail the shake-up specific correlation time delay $\tau_{e-e}$. \\
The interferometric RABBITT technique relies on the interference between two different two-photon pathways to the same continuum final state involving the absorption of one XUV photon from an attosecond pulse train (APT) and the absorption or stimulated emission of one IR photon with frequency $\omega_{\IR}$ from a time-delayed weak replica of the IR field which generated the APT.
The peaks in the photoelectron spectrum  reached via these two-photon transitions, commonly called sidebands, at energies $E$,
\begin{equation}
    P(E) \propto \left| \mathcal{A}^{(2)}_{f \leftarrow i} \left( E \right) \right|^2,
\end{equation}
show a characteristic oscillation as a function of the time delay $\Delta t$ between the APT and IR fields 
\begin{equation}\label{eq:2w_beat}
P \left(E\right) = A + B \cos\left(2\omega_{\IR} \Delta t - \Delta \Phi (E)\right)
\end{equation} 
with an energy dependent phase offset $\Delta \Phi (E)$.
For low IR intensities the oscillations of the sidebands as a function of $\Delta t$ are well described employing second-order perturbation theory \cite{VenTaiMaq1996,KluDahGis2011,DahLhuMaq2012,DahGueKlu2013}.
The functional form of Eq.~\eqref{eq:2w_beat} is independent of the emission angle $\theta$ of the photoelectron relative to the polarization direction of the colinear IR and APT field. 
The amplitude  $\mathcal{A}^{(2)}_{f \leftarrow i}$  for the transition to the final state  $\left| f  \right>$ corresponding to the sideband $H_{2n}$ can be described \cite{JimMarArg2016} as a superposition of two paths, namely absorption of one photon of the harmonic below the sideband ($H_{2n-1}$), followed by absorption of one photon of the fundamental IR field  $\mathcal{A}^{(2)}_{H_{2n-1}+\omega_{\IR}}$, and absorption of one photon of the harmonic above the sideband ($H_{2n+1}$) followed by stimulated emission of one IR photon $\mathcal{A}^{(2)}_{H_{2n+1}-\omega_{\IR}}$.
Accordingly, the transition amplitude is given by
\begin{equation} \label{eq:two_photon_ampl_2}
\mathcal{A}^{(2)}_{f \leftarrow i}\left(E=H_{2n}\right)=\mathcal{A}^{(2)}_{H_{2n-1}+\omega_{\IR}} +\mathcal{A}^{(2)}_{H_{2n+1}-\omega_{\IR}}.
\end{equation}
The phase offset of the $2 \omega_{\IR}$ beating in RABBITT traces, $\Delta \Phi$,  can be obtained from the phase difference between these two partial amplitudes
\begin{equation}\label{eq:phase_diff_rabbitt}
    \Delta \Phi \left( H_{2n} \right) =  \arg{\left[\mathcal{A}^{(2)}_{H_{2n+1}-\omega_{\IR}} \right]} - \arg{\left[\mathcal{A}^{(2)}_{H_{2n-1}+\omega_{\IR}} \right]} .
\end{equation}

\subsection{EWS and cc delays}
When each of the involved intermediate states ($H_{2n-1},\: H_{2n+1}$) and the final state involve structureless continuum states, experiment and theory \cite{KluDahGis2011,DahGueKlu2013,DahLhuMaq2012} have shown that the acquired phase $\Delta \Phi$ can be approximated as the sum of two distinct phases:
the one-photon half-scattering phase of the XUV triggered transition from the initial bound state  $\left| i  \right>$ to the intermediate continuum states $H_{2n-1}$ or $H_{2n+1}$, and the additional scattering phase acquired by the IR field induced continuum-continuum transitions ($H_{2n-1} \rightarrow H_{2n}$) or ($H_{2n+1} \rightarrow H_{2n}$).
In addition to these atomic phases, also phase differences between adjacent harmonic peaks of the APT may contribute \cite{VenTaiMaq1996,PauTomBre2001,Mul2002,MaiBohFra2003}.
As this XUV pulse related phase does not depend on the emission angle, we will omit this contribution in the following. \\
The additivity of the phase difference 
\begin{equation}
    \Delta \Phi \left( \theta \right) = \Phi \left(H_{2n+1}, \theta \right) - \Phi \left(H_{2n-1}, \theta \right),
\end{equation}
i.e.
\begin{equation}
    \Delta \Phi \left( \theta \right) = \Delta \Phi_{\EWS} \left( \theta \right) + \Delta \Phi_{\cc} \left( \theta \right) ,
\end{equation}
for any electron emission angle $\theta$ directly translates into the additivity of the corresponding angle-dependent time delays $\tau$ given by the finite-difference approximation to the spectral derivative
\begin{equation}\label{eq:rabbitt_delay}
    \tau (\theta) = \frac{\Delta \Phi (\theta)}{ 2 \omega_{\IR}}
\end{equation}
with
\begin{equation}\label{eq:tau_ews_plus_tau_cc}
    \tau (\theta) = \tau_{\EWS} (\theta)+\tau_{\cc} (\theta).
\end{equation}
It should be noted that while both time delays in Eq.~\eqref{eq:tau_ews_plus_tau_cc} are atomic EWS-type time delays for half-scattering, only the delay associated with the bound-continuum (bc) transition is conventionally referred to as EWS delay $\tau_{\EWS}=\tau_{\mathrm{bc}}$.
Recently \cite{FucDouDon2020} the delay associated with the continuum-continuum transition was also identified as an EWS-type delay.
For convenience, we adhere in Eq.~\eqref{eq:tau_ews_plus_tau_cc} to the standard convention.
The finite-difference approximation to the delay within RABBITT differs from the streaking protocol where the spectral derivative $\tau$ is directly observable.\\
The first term in Eq.~\eqref{eq:tau_ews_plus_tau_cc}, the one-photon bound-continuum delay 
\begin{equation}\label{eq:tau_ews_single_channel}
\tau_{\EWS}\left(\theta\right)=\frac{\partial}{\partial E} \arg \left\{ \left< \alpha,E, \theta \left| \hat{z}\right| i \right> \right\}= \frac{\partial}{\partial E} \eta_{\alpha}\left( E,\theta \right),
\end{equation}
is associated with the half scattering of the outgoing electron, transferred by the XUV pulse from the bound (ground) state $\left| i \right>$ to the (intermediate) continuum state $\left| \alpha, E, \theta \right>$ in the atomic potential.
Here,  $E$ and  $\theta$ denote the energy and emission angle of the departing electron, $\alpha$ comprises all other quantum numbers of the ion and electron, and $\eta_{\alpha}$ is commonly called scattering phase. \\
If the state $\alpha$ corresponds to a single partial wave with angular momentum $\ell$
(e.g. ionization by a dipole transition of an initial $s$ electron to a $p$ wave) $\tau_{\mathrm{EWS}}$ has no intrinsic angle dependence.
If, however, the intermediate state is a superposition of partial waves with different angular momenta (e.g. ionization of an electron out from the $2p_0$ state in neon to a superposition of an $s$ and $d_0$ wave) the scattering delay associated with the XUV transition itself becomes angle dependent \cite{WaeMosPav2014}.
For a statistical mixture of inital states (e.g. $2p_0$ and $2p_{m=\pm 1}$, of the $2p$ subshell), the angle dependent EWS delay is given by the ensemble average
\begin{equation}\label{eq:inc_sum_tau_ews}
\langle \tau_{\EWS} \rangle_{\alpha} \left( \theta \right)=\frac{\sum_{i}  \sigma_{\alpha_i}\left( \theta \right)\tau_{\EWS}^{(\alpha_i)}\left( \theta \right)} {\sum_{i}   \sigma_{\alpha_i}\left( \theta \right)},
\end{equation}
weighted by the cross sections $\sigma_{\alpha_i}\left( \theta \right)=\left| \left< \alpha_i, \varepsilon, \theta \left| \hat{z}\right| i \right>  \right|^2$.
\\
The second contribution to Eq.~\eqref{eq:tau_ews_plus_tau_cc}, the continuum-continuum delay $\tau_{\cc}$, is caused by the transition between continuum states in the atomic potential induced by absorption or emission of an IR photon \cite{KluDahGis2011}.
Unlike for streaking where the IR field employed is, typically, stronger, the continuum-continuum transition in the RABBITT protocol is well described by lowest-order perturbation theory.
Employing an asymptotic expansion for large $kr$ of the outgoing Coulomb wave and neglecting partial wave interferences, a simplified analytic asymptotic expression for the phase shift $\phi_{\cc}^{\mathrm{asym}}$ and, in turn, $\tau_{\cc}^{\mathrm{asym}}$ has been derived \cite{DahLhuMaq2012}
\begin{align}\label{eq:tau_cc_pt}
\tau_{\cc}^{\mathrm{asym}} &  \left(E,\omega_{\IR} \right)= \\
&\frac{ \phi^{\mathrm{asym}}_{\cc}\left( E, E + \omega_{\IR} \right) - \phi^{\mathrm{asym}}_{\cc}\left( E, E - \omega_{\IR} \right)}{2 \omega_{\IR}}.  \nonumber 
\end{align}
In this limit $\tau_{\cc}^{\mathrm{asym}}$ is independent of the emission angle. 
Only when numerically including non-asymptotic contributions from smaller values $kr$ of the Coulomb wave to the cc transition matrix elements and partial wave interferences, a $\theta$-dependent $\tau_{\cc}\left(E,\omega_{\IR},\theta  \right)$ emerges \cite{DahGueKlu2013,CirMarHeu2018}.

\subsection{Two-electron delay for correlation satellites}

An additional contribution to the phase shift of the wavepacket and, thus, to the time delay originates from true two-electron processes beyond the direct ionization discussed above.
A prototypical case is the photoionization of helium accompanied by shake-up of the residual electron,
\begin{equation} \label{eq:shake_up}
    \hbar \omega + \mathrm{He}(1s^2) \rightarrow \mathrm{He}^+(n\ell m) + e,
\end{equation}
often referred to as correlation-satellite lines in the photoelectron spectrum.
For this process, the two-electron wavepacket acquires an additional dynamical phase and time delay within a streaking or RABBITT protocol, beyond the bound-continuum EWS and the cc contribution.
The IR field polarizes the (quasi) degenerate $n$ manifold of the bound electron formed in the correlated ionization process, and the resulting dynamical Stark shift imprints an additional phase on the two-electron wave function which, in turn, contributes to the scattering phase of the entangled photoelectron. 
The resulting two-electron time delay can be analytically estimated as \cite{PazFeiNag2012,OssSieShi2017}
\begin{equation}\label{eq:dipole_delay}
\begin{split}
\tau_{e-e}\left(\theta\right) & =\frac{1}{\omega_{\mathrm{IR}}} \mathrm{atan}\left(-\omega_{\mathrm{IR}}\frac{ \textbf{d} \cdot \bm{\hat{\pi}_{\IR}}}{\textbf{k} \cdot \bm{ \hat{\pi}_{\IR}}}\right) \\
&=\frac{1}{\omega_{\mathrm{IR}}} \mathrm{atan} \left(-\omega_{\mathrm{IR}}\frac{ d_z }{k\cos \left(\theta\right)} \right),
\end{split}
\end{equation}
where $\bm{d}$ is the dipole moment of the bound electron, $\bm{\hat{\pi}_{\IR}}$ is the polarization direction of the IR field, $\textbf{k}$ is the momentum of the outgoing electron, and we have assumed that the IR field is linearly polarized along the $\hat{z}$ direction.
A similar contribution to the time delay has originally been predicted for molecules with a permanent dipole moment \cite{BagMad2010}.
For streaking experiments, where the emitted photoelectrons are only measured in forward direction, the analytic estimate has been shown to be accurate for the prototypical example of shake-up ionization in helium \cite{PazFeiNag2012,OssSieShi2017}. \\
Eq.~\eqref{eq:dipole_delay} extends this estimate to angles $\theta \neq 0^{\circ}$ which become accessible by angle-resolved RABBITT.
Assuming that the additivity of the time delay holds also for $\tau_{e-e}$, the total angle-dependent time delay $\tau^{(t)} (\theta)$ is given by
\begin{equation}\label{eq:time_delay}
    \tau^{(t)} (\theta) = \tau_{\EWS}(\theta)+\tau_{\cc}(\theta)+\tau_{e-e}(\theta).
\end{equation}
In the following we will numerically test Eq.~\eqref{eq:time_delay}. 
Probing $\tau_{e-e}(\theta)$ by RABBITT is of particular importance as the resolution of different correlation satellite lines with increasing $n$ requires high spectral resolution as achieved by such a protocol \cite{IsiSquBus2017} but is difficult to realize by an attosecond streaking camera.


\section{Numerical results for the angular variation of the time delay}\label{sec:angle_dependence_tau_r}

We explore in the following the relative importance of the different contributions to the total time delay [Eq.~\eqref{eq:time_delay}] with the help of numerical results for a few prototypical cases before presenting detailed results for the helium shake-up satellites.

\subsection{Angle dependence of the EWS delay}

\begin{figure}[tb]
    \includegraphics[width=1.\columnwidth]{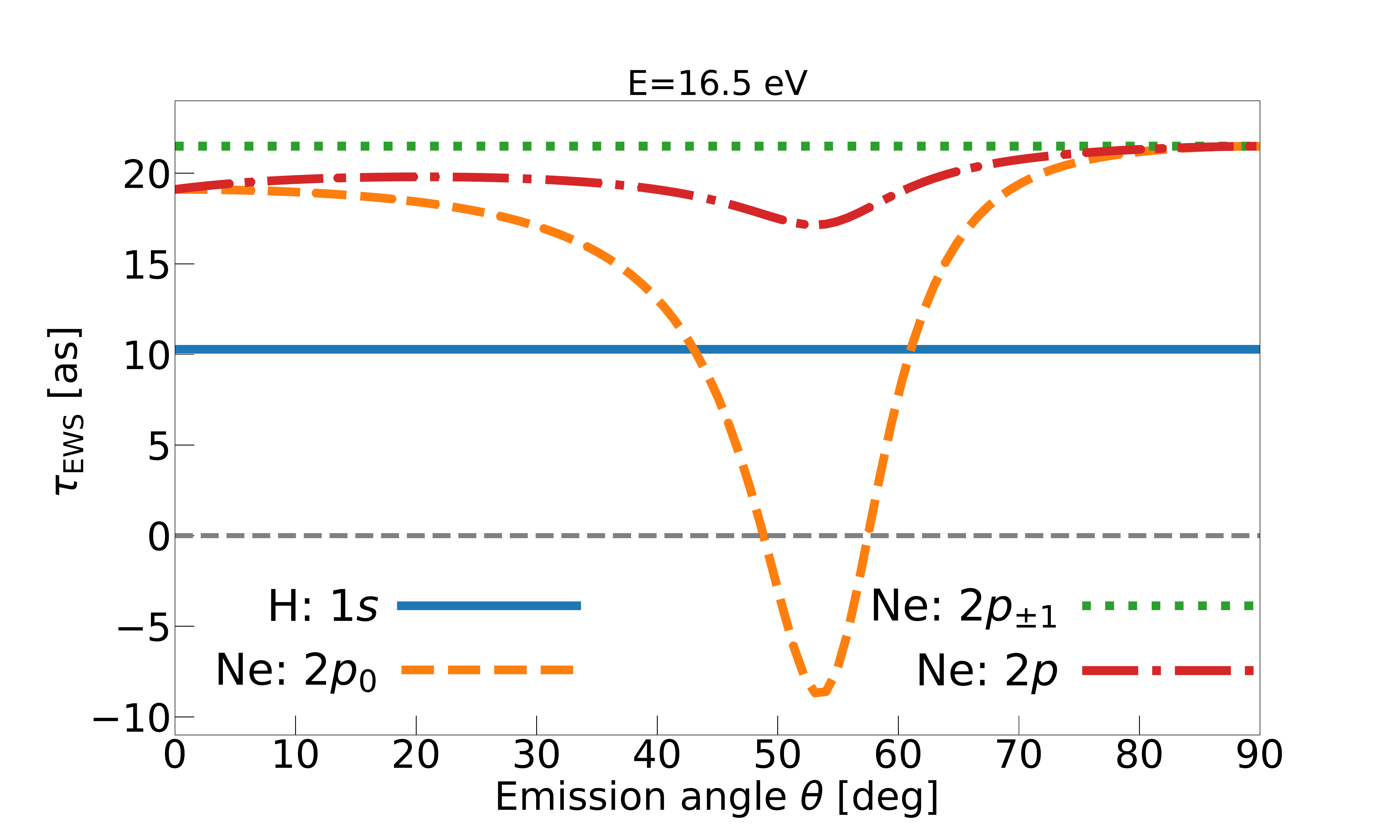}
	\caption{Bound-free scattering time delays $\tau_{\mathrm{EWS}}$ for a final photoelectron energy of 16.5~eV computed numerically for ionization of the $1s$ ground state of hydrogen and the $2p_{m=0}$ and the $2p_{m=\pm1}$ states of neon represented by a model potential \cite{TonLin2005} yielding the experimental ionization energy.
	The subshell average $\left< \tau_{\mathrm{EWS}}\right>_{2p}$ is obtained using Eq.~\eqref{eq:inc_sum_tau_ews}.} \label{fig:ews_h_ne}
\end{figure}

The partial-wave interference as origin of the angle dependence can be demonstrated for prototypical cases, the ionization of the $1s$ state of hydrogen and of the $2p_0$ and $2p_{\pm1}$ states of neon described by a model potential \cite{TonLin2005} (Fig.~\ref{fig:ews_h_ne}).
For the hydrogen ground state, the outgoing electron wavepacket created by the absorption of a linearly polarized XUV photon is formed solely by a $p_{0}$ wave, $\left| E, \ell=1, m=0\right>$, resulting in an angle-independent delay.
By contrast, photoionization of the $\left|2 p,m=0 \right>$ state of neon creates a wavepacket containing a superposition of an $s$-wave, $\left| E, \ell=0, m=0\right>$, and a $d_0$-wave, $\left| E, \ell=2, m=0\right>$.
Cross sections and spectral phases of these two partial waves are different and, thus, $\tau_{\mathrm{EWS}}$  shows a characteristic dependence on $\theta$ \cite{WaeMosPav2014}.
Photoionization of the Ne $\left| 2p, m = \pm 1 \right>$ states, on the other hand, gives access only to a single partial wave $\left| E, \ell=2, m=\pm 1\right>$.
Consequently, $\tau_{\EWS}$ is again independent of $\theta$.
The ensemble average $\left<  \tau_{\EWS} \right>_{2p}$ over the $2p$ subshell yields only a weak $\theta$ dependence (Fig.~\ref{fig:ews_h_ne}). 
This is due to the fact that the strong variation with angle of $\tau_{\EWS}$ for the $2p_0$ state is confined to the region where the cross section is small and where ionization from the $2p_{\pm 1}$ states dominates.
Overall, the observed angular variation of $\tau_{\EWS}$ is rather small.
Further, for large electron emission angles it is orders of magnitude smaller than the angular variation in RABBITT experiments \cite{HeuJimCir2016,CirMarHeu2018} and simulations \cite{IvaKhe2017,BraNasKhe2018} in this case.
It is also much smaller than the variation of the two-electron delay $\tau_{e-e}(\theta)$ discussed in the following.


\subsection{Angle dependence of the continuum-continuum delay}\label{sec:angle_dependence_tau_cc}
\begin{figure}[tbp]
\begin{center}
\includegraphics[width=1\columnwidth]{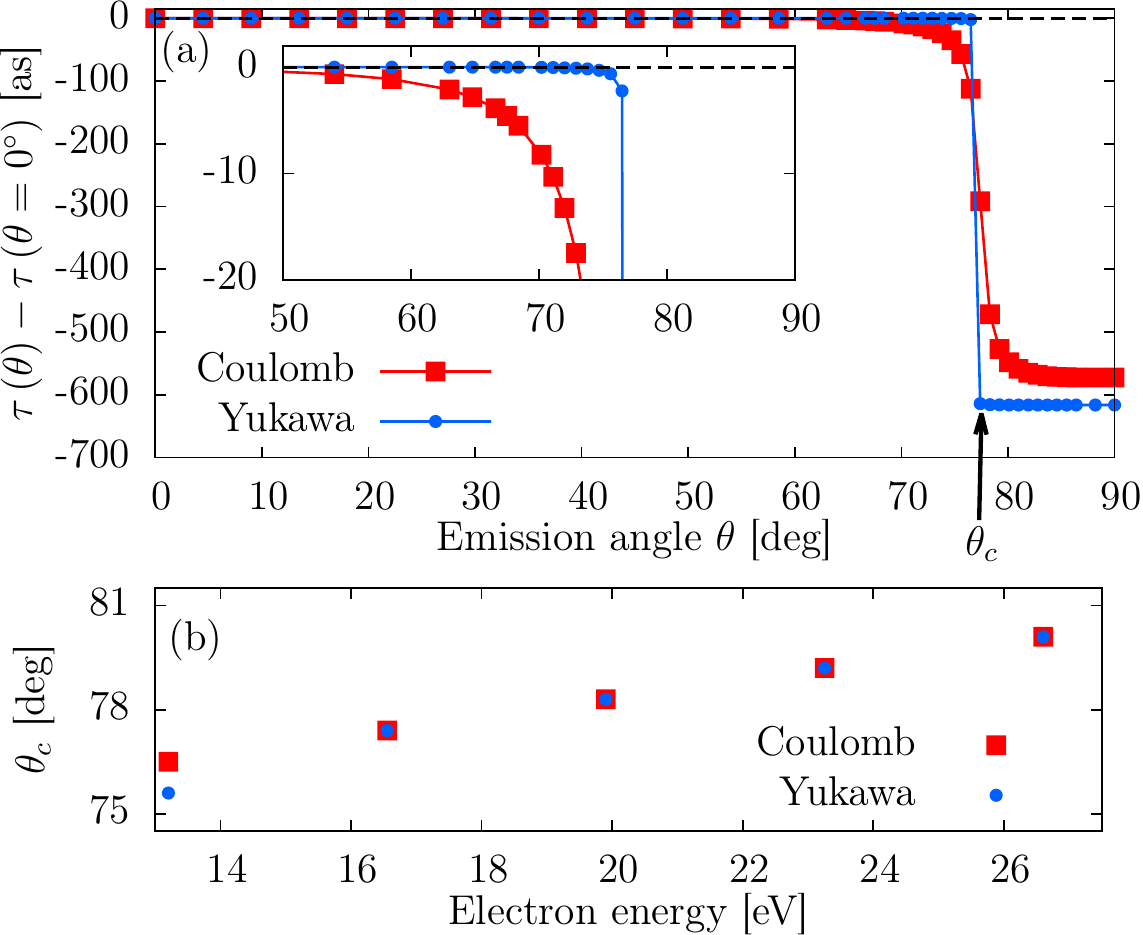}
\end{center}
\caption{(a) Angle dependence of the relative time delay $\tau(\theta)-\tau(\theta=0^{\circ})$, at the sideband centered at $E=$16.5~eV for the Yukawa potential (blue circles) and hydrogen (red squares).
(b) Critical angle $\theta_c$ at which the jump in $\Delta \Phi$ and $\tau$ occurs, shown for the Yukawa potential and hydrogen.
The wavelength of the IR is $\lambda_{\IR}=740$~nm.
}\label{fig:yuk_angle}
\end{figure}
The angle dependence of the continuum-continuum delay $\tau_{\cc}$ was recently observed for the direct ionization of helium \cite{HeuJimCir2016} and argon \cite{BusVinZho2018,CirMarHeu2018}. 
Moreover, also the angular momentum dependence of the phase acquired by the IR photon absorption or emission in the continuum-continuum transition was analyzed \cite{CirMarHeu2018,FucDouDon2020}.
In order to disentangle the influence of long-range Coulomb interaction from partial-wave interference effects contributing to the $\theta$ dependence of $\tau_{\cc}$ we compare RABBITT simulations for the $1s$ state of hydrogen with that for a model atom bound by a short-ranged Yukawa-like potential 
\begin{equation}
    V_{\mathrm{Y}}(r)=-\frac{1.90831}{r}e^{-r},
\end{equation}
with the same ground-state binding energy $E_{1s}=-0.5$~a.u. and $\ell=0$.
The simulation employs the pseudo-spectral method \cite{TonChu1997,NagPazFei2011} (see appendix for numerical details).
Previous simulations for a streaking setting \cite{NagPazFei2011} have shown that for short-ranged potentials the total time delay is given by $\tau_{\EWS}$, i.e. the cc contribution is negligible.
Differently, in the presence of long-ranged interaction the coupling of the IR field to the Coulomb continuum gives an additional Coulomb-laser contribtion $\tau_{\mathrm{CLC}}$ which has been found to closely agreee with $\tau_{\cc}$ from RABBITT settings \cite{PazNagBur2015,Don2019}. \\
We explore now the $\theta$ dependence of $\tau_{\cc}$ in a RABBITT setting for a long-ranged and short-ranged potential in photoionization by an APT of the odd harmonics from 15 to 31 of an IR field with $\lambda_{\IR}=740$~nm.
Focusing on the sideband centered at 16.5~eV (the other sidebands show a very similar behavior), the time delay obtained for the Yukawa potential (Fig.~\ref{fig:yuk_angle}a) shows almost no angle dependence up tp $\theta \approx$~75$^{\circ}$ where a phase jump of $\pi$ occurs [corresponding to a time delay change of approximately 617~as for this specific IR wavelength, see Eq.~\eqref{eq:rabbitt_delay}].
The time delay for hydrogen closely mimics this behavior, however, smooths the phase jump and reduces the jump height from $\pi$ to 0.92$\pi$.
The reduction in jump height can be traced to the angular momentum dependence of $\phi_{\cc}$ \cite{DahGueKlu2013} and to the propensity rule that for photoabsorption the increase while for photoemission the decrease in angular momentum is preferred \cite{Fan1985,BetRotBel1980}.
The critical angle where the jump occurs, $\theta_c\simeq$~75$^{\circ}$, is identical for both potentials.
 \\
In order to identify the origin of this rapid phase jump, we employ second-order perturbation theory for absorption of one XUV photon and subsequent absorption (emission) of an IR photon to the wavepacket following   \cite{DahLhuMaq2012,BusVinZho2018} for the short-ranged potential
\begin{align}\label{eq:model_jump_angle_psi_yuk}
\Psi &\left(E,\Delta t,\theta\right) =  \sum_{L=0,2}e^{i\frac{\pi}{2 }}Y_L^0 \left(\theta\right) \nonumber \\
& \left\{ \left|\mathcal{A}_{pL}^{(+)}\right|e^{i \left[ \omega_{\mathrm{IR}} \Delta t +  \eta_p^{(-)} \right]} +\left|\mathcal{A}_{pL}^{(-)}\right|e^{i \left[- \omega_{\mathrm{IR}} \Delta t +  \eta_p^{(+)} \right]}   \right\}.
\end{align}
The two-photon transition amplitude for absorption of one XUV photon to a $p$ wave (angular momentum $\ell=1$) and subsequent absorption ($\mathcal{A}_{pL}^{(+)}$) or emission ($\mathcal{A}_{pL}^{(-)}$) of an IR photon  to the final angular momentum $L$, $\mathcal{A}_{pL}^{(\pm)}$, is assumed to have  three phase contributions:
the phase due to half-scattering at the centrifugal potential, $\ell \pi /2$, the half-scattering phase due to the ``atomic'' potential $\eta_{\ell}^{(\mp)}=\eta_{\ell}\left(E \mp \omega_{\IR} \right)$, and the additional phase due to the XUV-IR pump probe delay $\pm \omega_{\IR} \Delta t$.
For a short-ranged ``atomic'' potential, the IR induced continuum-continuum transition of the outgoing wavepacket does not generate a significant additional phaseshift, unlike the $\phi_{\cc}$ phase in the Coulomb potential. 
\newline
\begin{figure}[tbp]

\includegraphics[width=1\columnwidth]{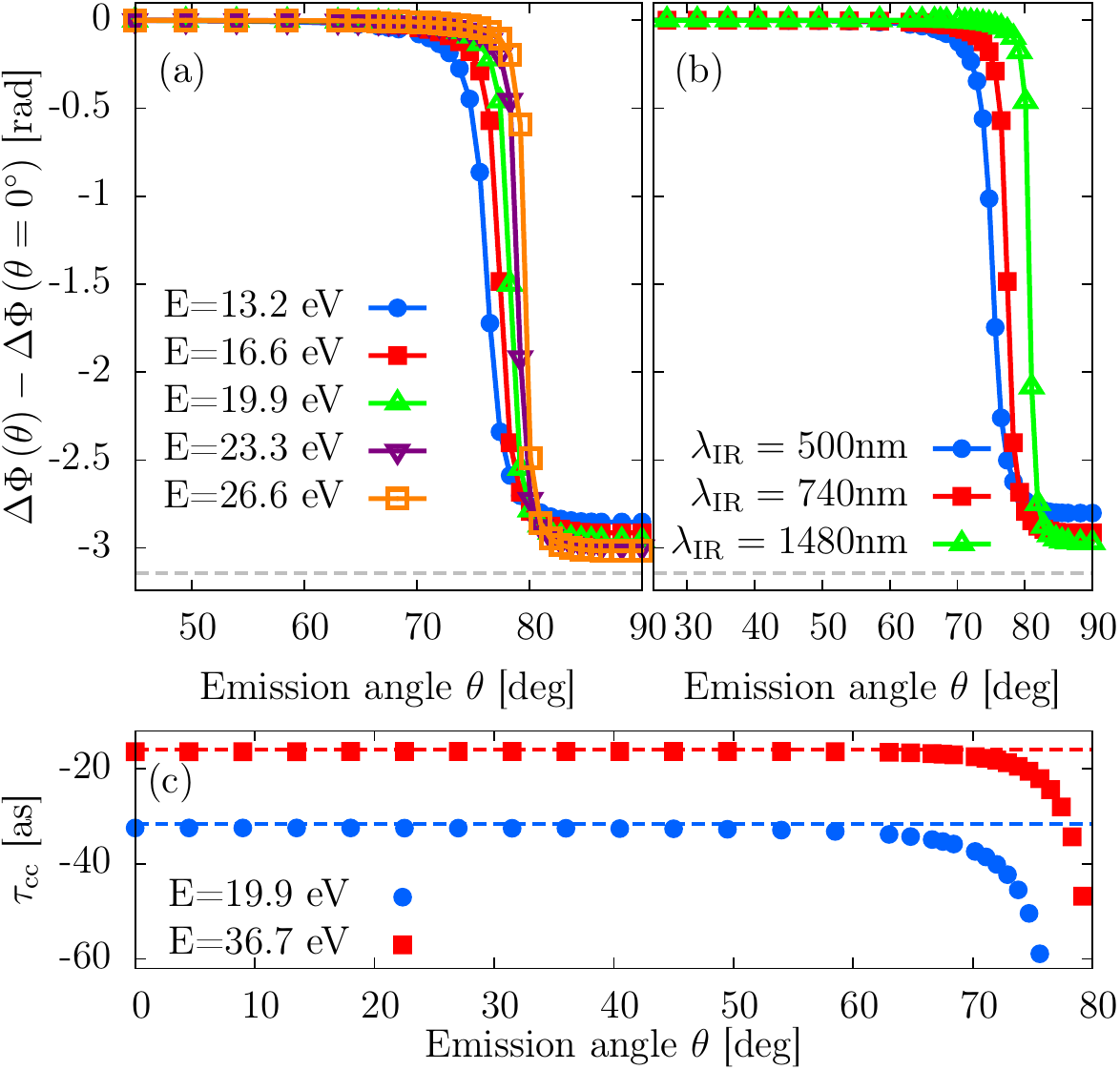}
\caption{
Angle dependence of the RABBITT phase $\Delta \Phi$  for (a) different sideband energies $E$ obtained for ionization of hydrogen ($\lambda_{\IR}=$~740~nm), (b) three different APTs and fundamental IR fields.
The sideband energies are 16.15~eV ($\lambda_{\IR}=$~500~nm) and 16.5~eV ($\lambda_{\IR}=$~740~nm, and 1480~nm). \\
(c) Angle dependence of $\tau_{\cc}$ for ionization of hydrogen comparing the asymptotic prediction [Eq.~\eqref{eq:tau_cc_pt}] (dashed) and numerical evaluation (symbols) for two energies and $\lambda_{\IR}=$~740~nm (for details see text). 
}\label{fig:angle_wavelength}
\end{figure}
The angular resolved photoelectron spectrum for a given sideband with energy $E=H_{2n}$ is proportional to
\begin{align}\label{eq:osc}
P&\left(E=H_{2n},\theta \right) \propto \left|\Psi\left(E,\Delta t,\theta\right)\right|^2 \\
&= A\left(E,\theta\right)+ 2  \left|\mathcal{A}_{p0}^{(+)}\right|^2 B\left(E,\theta\right)  \cos\left(2 \omega_{\IR} \Delta t - \Delta \eta_p \right), \nonumber
\end{align}
with $\Delta \eta_p  =  \left[ \eta_p \left( E + \omega_{\IR}\right)-\eta_p \left( E - \omega_{\IR}\right) \right] $, and
\begin{align}
    A \left(E,\theta \right)  &=  Y_0^0\left(\theta\right)^2 \left[ \left|\mathcal{A}_{p0}^{(+)}\right|^2 + \left|\mathcal{A}_{p0}^{(-)}\right|^2 \right]  \\
    & + Y_2^0\left(\theta\right)^2 \left[ \left|\mathcal{A}_{p2}^{(+)}\right|^2  +\left|\mathcal{A}_{p2}^{(-)}\right|^2 \right] \nonumber \\ 
    &+ 2 Y_0^0\left(\theta\right) Y_2^0\left(\theta\right) \left[ \left|\mathcal{A}_{p0}^{(+)}\right| \left|\mathcal{A}_{p2}^{(+)}\right|+\left|\mathcal{A}_{p0}^{(-)}\right|\left|\mathcal{A}_{p2}^{(-)}\right| \right] \nonumber
\end{align}
is independent of the XUV-IR delay $\Delta$t.
The amplitude of the oscillation $\sim \cos \left( 2 \omega_{\IR} \Delta t - \Delta \eta_{p}  \right)$ is governed by
\begin{equation}\label{eq:b}
B\left(E,\theta\right)= a Y_0^0\left(\theta \right)^2+b c^2 Y_2^0\left(\theta\right)^2 + Y_0^0\left(\theta\right) Y_2^0\left(\theta\right)c  \left(1+ab\right)
\end{equation}
with
\begin{align}
a= \left|\mathcal{A}_{p0}^{(-)}\right| / \left|\mathcal{A}_{p0}^{(+)}\right|,  \label{eq:par_a}\\
b= \left|\mathcal{A}_{p2}^{(+)}\right| / \left|\mathcal{A}_{p2}^{(-)}\right|, \\
c=\left|\mathcal{A}_{p2}^{(-)}\right|/\left|\mathcal{A}_{p0}^{(+)}\right|. \label{eq:par_c}
\end{align}
As $Y_2^{0}(\theta)$ changes sign at the ``magic'' angle  $\theta= 54.7^{\circ}$, $B(E,\theta)$ may eventually change sign at the critical angle  $\theta_c > 54.7^{\circ}$ given by the condition
\begin{equation}
     a Y_0^0\left(\theta_c \right)^2+b c^2 Y_2^0\left(\theta_c\right)^2 = c  \left(1+ab\right) Y_0^0\left(\theta_c\right) \left| Y_2^0\left(\theta_c\right)\right|.
\end{equation}
This sign change results in a phase jump of the retrieved phase $\Delta \Phi$, observed in Fig.~\ref{fig:yuk_angle}a for the Yukawa potential.
The angle at which $B\left(E, \theta\right)$ changes sign, varies with the final photoelectron energy, and depends on the parameters $a,b,c$ [Eq.~\eqref{eq:par_a}--Eq.~\eqref{eq:par_c}].
\newline
\begin{figure}[tbp]

\includegraphics[width=1\columnwidth]{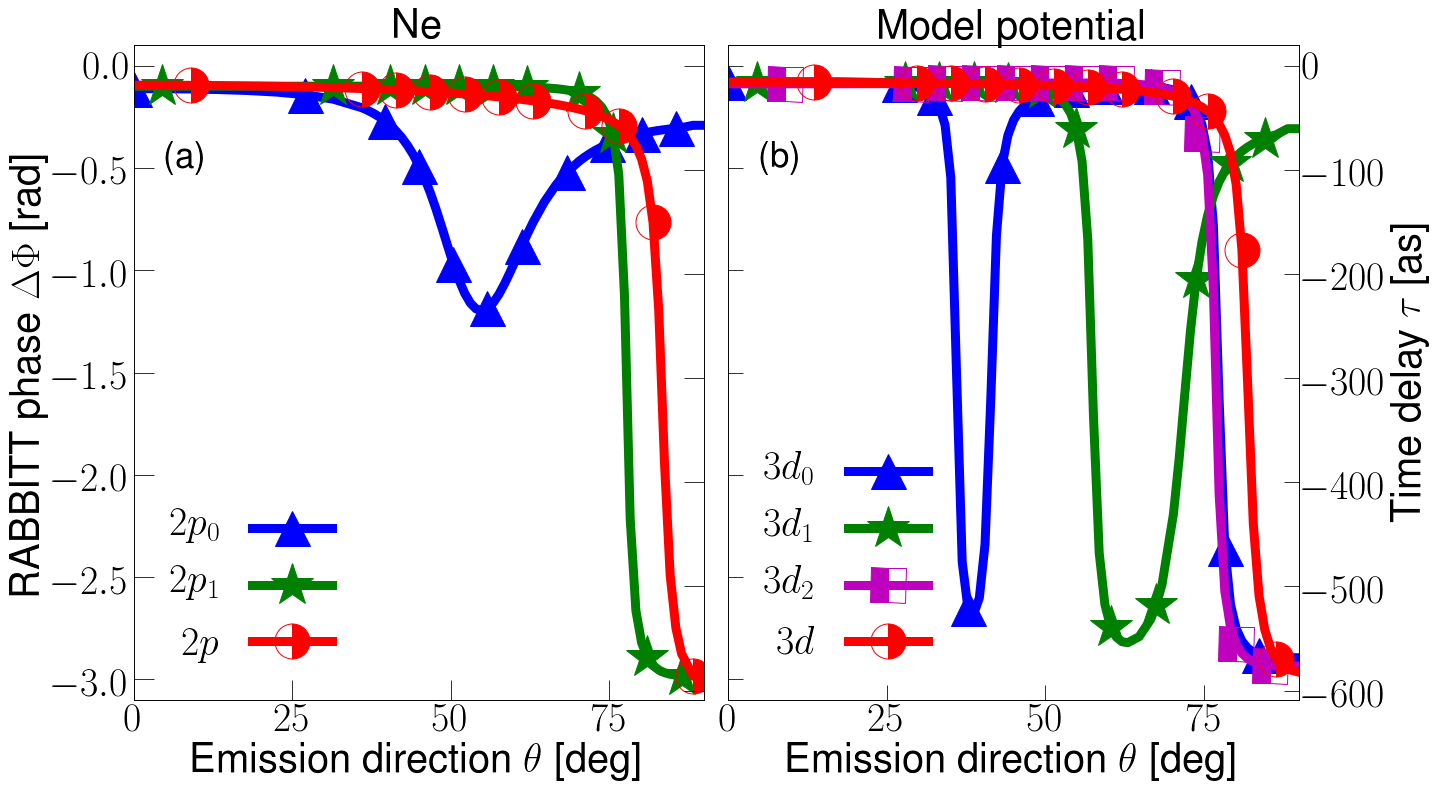}
\caption{
Angular-resolved RABBITT phase $\Delta \Phi$  obtained for ionization from different initial states:
(a) $2p_m$ states of neon described by a model potential \cite{TonLin2005}, and (b) $3d_m$ states bound by a  model potential $V=-\left(1 + 10 e^{-16 r} + 14 e^{-1.2 r} \right)/r $.
The final electron energy in the  sideband considered is 18.8~eV in (a) and 18.7~eV in (b), and
$\lambda_{\IR}$ =740~nm.}\label{fig:phase_neon_d_pot}
\end{figure}
Our simulations show that the critical angle increases monotonically with the sideband energy (Fig.~\ref{fig:yuk_angle}b) which is in line with previous results for helium \cite{HeuJimCir2016,IvaKhe2017}, neon \cite{IvaKhe2017}, and argon \cite{BusVinZho2018,CirMarHeu2018}.
Further, the height of the phase jump for hydrogen, which is smaller than $\pi$ due to the dependence of $\phi_{\cc}$ on the intermediate and final angular momenta, approaches asymptotically $\pi$ as the energy of the photoelectron is increased (Fig.~\ref{fig:angle_wavelength}a).
Similarly, the phase jump approaches $\pi$ when increasing the wavelength of the IR field while keeping the sideband energy fixed (Fig.~\ref{fig:angle_wavelength}b).
This is, again, an effect of the angular momentum dependence of the $\phi_{\cc}$ phase in the Coulomb field. 
Further, the emission angle at which the phase jump occurs increases with the IR wavelength, which is due to the dependence of $\left|\mathcal{A}_{pL}^{\left(\pm \right)}\right|$ on the IR wavelength.
Our numerical results for hydrogen allow the determination of the angle dependence of $\tau_{\cc}$ by applying Eq.~\eqref{eq:tau_ews_plus_tau_cc}, i.e. subtracting $\tau_{\EWS}$ from the simulation results to obtain $\tau_{\cc}(\theta)$ (Fig.~\ref{fig:angle_wavelength}c). 
Comparing these numerical results to the asymptotic prediction $\tau_{\cc}^{\mathrm{asym}}$ [Eq.~\eqref{eq:tau_cc_pt}], we find almost perfect agreement for $\theta<60^{\circ}$.
The rapid jump observed for higher emission angles is not reproduced by the asymptotic expansion.
As expected we find that for increasing electron energy, the asymptotic  $\tau_{\cc}^{\mathrm{asym}}$ agrees with the simulation results up to larger emission angles.
\newline
As the appearance of the phase jump is a direct consequence of the IR field induced partial wave interference, its position and shape is sensitively dependent on the angular momentum of the initial bound state to be ionized.
Starting from different initial $\ell$, the two-photon process gives access to different superpositions of partial waves.
This is illustrated for ionization from the $2p$ shell of neon, and from the $3d$ shell bound by a model potential  $V=-\left(1 + 10 e^{-16 r} + 14 e^{-1.2 r} \right) /r$ (Fig.~\ref{fig:phase_neon_d_pot}).
In all cases, a single-active electron approximation is employed.
A simple systematic pattern emerges: all initial states with the largest angular momentum quantum number of a given shell $\ell=n-1$, $\left| m \right|=n-1$ (e.g. $1s$, $2p_{\pm 1}$, $3d_{\pm 2}$) display a phase jump at a critical angle $\theta_c \geq 75^{\circ}$.
In each case, only two partial waves are accessible by the final IR transition with only one of which featuring a spherical harmonic with a node for $\theta \in \left( 0^{\circ},90^{\circ} \right)$, i.e. $\left( Y_{2}^0, Y_{3}^1, Y_{4}^2  \right)$, giving rise to a change of sign.
A qualitatively different shape appears for those initial states (e.g. $2p_0$, $3d_{\pm 1}$) which give rise to an interference among three partial waves.
Here the phase increases by almost $\pi$ at angles $\theta \leq 60^{\circ}$ before reverting back close to zero near $90^{\circ}$.
For initial states where even more interfering pathways lead to the same final state (e.g. four independent pathways to three partial waves for ionization of $3d_0$) the peak in the phase excursion moves to much smaller $\theta$ ($\simeq 40^{\circ}$).
Finally, after performing the subshell average $\left< \Delta \Phi \right>_{n \ell}$ over all $m$ states, the angular variation of the phase change reduces for all $n \ell$ to the simple phase jump of the node-free initial $1s$.
This is a consequence of the small cross sections in the range of large phase excursions.
These qualitative features are governed by the partial-wave distribution in the continuum final state and are only marginally affected by the long-range tail of the Coulomb field.
This non-Coulombic contribution to the continuum-continuum phase of RABBITT is key to understand the dependence on emission angle. 
\begin{figure}[tb]
\centering
\includegraphics[width=1\columnwidth]{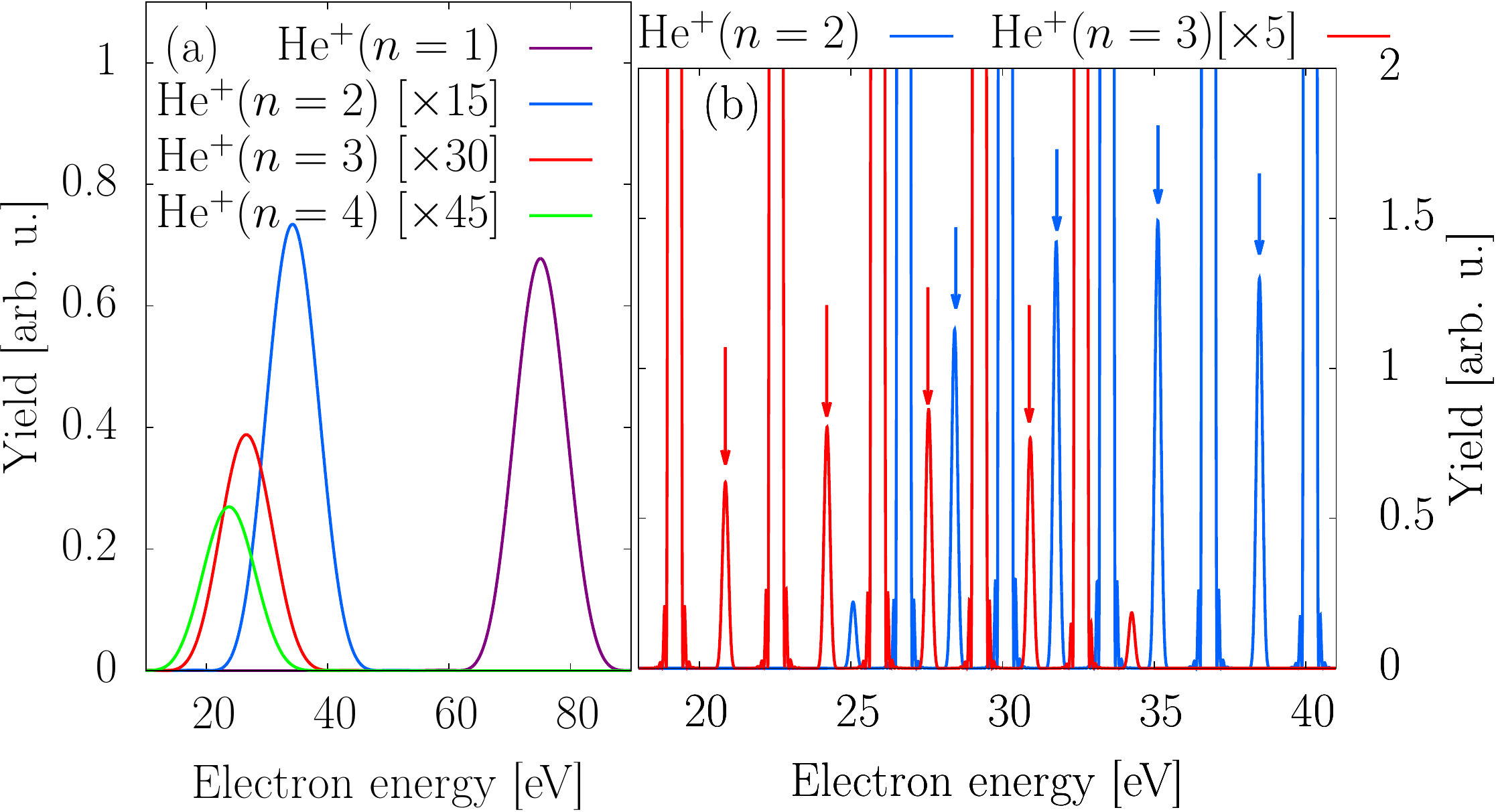}
\caption{
Electron spectra for different He$^+(n)$ correlation satellites for photoionization of helium by (a) a 300 as FHWM duration XUV pulse with $E_{\XUV}=$~100~eV.
(b) RABBITT spectrum for the $n=2$ (blue) and $n=3$ (red) channels and $\Delta t=0$.
The high spectral resolution provided by RABBITT allows to spectrally select the  sidebands (marked by arrows).
The spectra for the different channels are scaled to enhance visibility.
} \label{fig:shake_up_trace_740nm}
\end{figure}
\subsection{Angle dependence of the two-electron delay for shake-up ionization in helium}\label{sec:shake_up_angle}
We turn now to the two-electron induced delay for photoionization of helium accompanied by shake-up excitation of the bound electron [Eq.~\eqref{eq:shake_up}].
For electron emission along the polarization direction, the time delay for these correlation satellites has been theoretically \cite{PazFeiNag2012} and experimentally investigated \cite{OssSieShi2017}.
Here we present the first study of the angular variation of the time delay for correlation satellites.
We compare and contrast the behavior of the $\theta$ dependence of $\tau_{e-e}$ with that of $\tau_{\EWS}$ and $\tau_{\cc}$. 
\begin{figure}[tb]
\centering
\includegraphics[width=1\columnwidth]{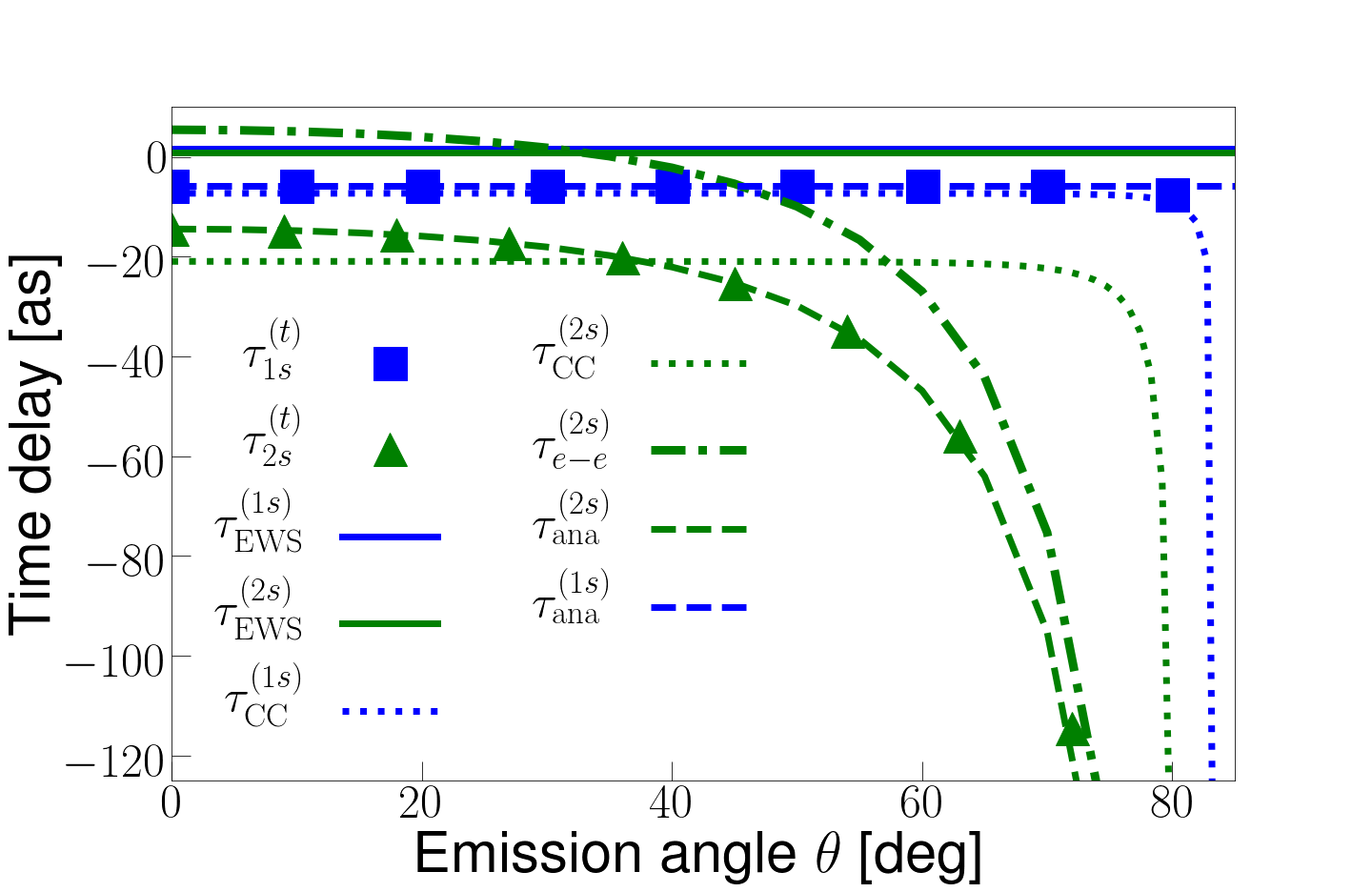}
\caption{
Angle dependence of the time delay of ($1s$) direct ionization (blue) and for the $2s$ correlation satellite (green). 
Shown are the total time delays $\tau_{1s}^{(t)}$ (blue squares) and $\tau_{2s}^{(t)}$ (green triangles) obtained from the simulations, as well as their EWS (solid lines), cc (dotted lines), and e-e contributions (dashed-dotted line). For the analytic prediction $\tau_{\mathrm{ana}}$ (dashed lines) using the additivity rule [Eq.~\eqref{eq:time_delay}] we take $\tau_{\cc}^{\mathrm{asym}}$ [Eq.~\eqref{eq:tau_cc_pt}].
$\tau_{\cc}(\theta)$ is extracted from RABBITT simulations of helium in the single-active electron approximation for final electron energies of 69.2~eV (1s) and 28.4~eV (2s).
} \label{fig:jb}
\end{figure}
\newline
During shake-up ionization of a multi-electron system the ionized electron interacts with the residual electron via electron-electron interaction and thus can promote excitations of the ion.
Consequently, the kinetic energy of the emitted electron is reduced compared to the direct ionization where the residual ion stays in the ground state (``main line'').
For the prototypical system of helium this leads to the well-known correlation satellites in the photoelectron spectrum at energies
\begin{align}
E_{n}&=   E_{\XUV}-I_{p,1} - \left(E_{\mathrm{He}^+(n)}-E_{\mathrm{He}^+(1)} \right)          \\
&=E_{\XUV}-I_{p,1} - 2 \frac{n^2-1}{n^2}.  \nonumber
\end{align}
$ E_{\XUV}$ is the energy of the absorbed XUV photon, $I_{p,1}$ is the first ionization potential for helium (0.904~a.u.), and $E_{\mathrm{He}^+(n)}=-\frac{2}{n^2}$~a.u. is the energy of the He$^+(n)$ residual ion. 
If the spectral width of the ionizing pulse is narrow enough the peaks do not overlap  (Fig.~\ref{fig:shake_up_trace_740nm}b).
The spectral overlap between different shake-up channels, however, is an inherent challenge to attosecond streaking because two effects limit spectral sensitivity.
The spectral width of the single attosecond pulse of a few hundred attoseconds duration used in the streaking protocol is, generally, too broad to resolve different satellites for  $n \geq 2$ from each other.
This can be seen in Fig.~\ref{fig:shake_up_trace_740nm}a where we use an XUV with full-width at half maximum (FHWM) duration of 300~as typically employed in streaking \cite{OssSieShi2017}.
Moreover, the amplitude of the oscillatory energy shift $E(\Delta t)=E_n+\sqrt{2 E_n}A_{\IR} (\Delta t)$ by the moderately strong IR streaking field may become comparable to the spacing between adjacent satellites.
This  renders the analysis of streaking spectra in the presence of overlapping shake-up channels quite difficult as was seen, e.g., for photoionization of neon \cite{SchFieKar2010,FeiZatNag2014}.
The RABBITT protocol offering simultaneously time- and energy resolution \cite{IsiSquBus2017} promises improved access to time delay information of correlation satellites.
To analyze the shake-up delays for atomic helium accessible by RABBITT we employ time-dependent \textit{ab initio} simulations \cite{FeiNagPaz2008,DonBreNi2019}.
We choose an IR pulse with wavelength $\lambda_{\mathrm{IR}}=$~740~nm, and an APT consisting of the 55$^{\mathrm{th}}$, 57$^{\mathrm{th}}$, 59$^{\mathrm{th}}$, 61$^{\mathrm{st}}$, and 63$^{\mathrm{rd}}$ harmonic.
For more details see appendix.
Due to the high spectral resolution of RABBITT the electron spectra associated with the main line $(n=1)$ for direct ionization and the correlation satellites $(n=2,3)$ are well resolved.
\\
For direct ionization the photoionization time delays are in very good agreement with single-active electron calculations showing that electronic correlations do not have a significant influence on the  photoionization time delay in this case \cite{Paz2013,PazNagBur2015,OssSieShi2017}.
\begin{figure}[tb]
\centering
\includegraphics[width=1\columnwidth]{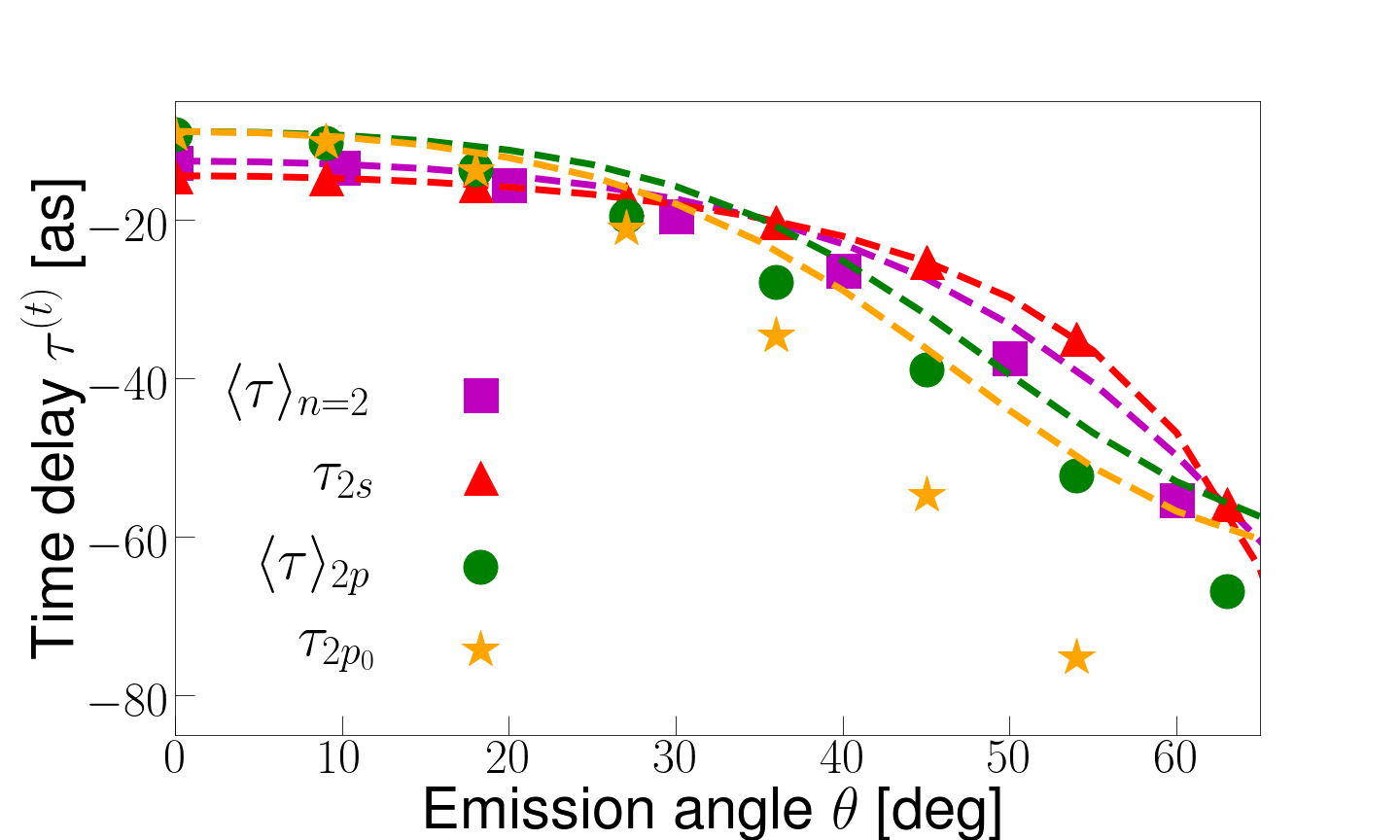}
\caption{
Time delays for the $n=2$ shake-up channels for the lowest energetic sideband (E=28.4~eV).
$\left< \tau \right>_{n=2} \left(  \theta \right)$ for the full $n=2$ shell, and resolved for the angular momentum eigenstates of the ion $2s$ and $2p_0$. Electron correlation effects also populate the ionic states  $2p_{\pm 1}$ contributing to the delay of the $2p$ subshell $\left< \tau \right>_{2p}$.
Dashed lines represent the result from Eq.~\eqref{eq:time_delay}.
} \label{fig:he_n2_shakeups}
\end{figure}
For shake-up channels a drastically different picture emerges (Fig.~\ref{fig:jb}).
Here a single-active electron approximation fails and $\tau_{e-e}$ plays an important role.
The total time delay $\tau^{(t)}_{2s}$ becomes strongly angle dependent for the shake-up state at angles well below the critical angle $\theta_c$ where the partial wave interference induced phase jump occurs.
For the $2s$ state the critical angle $\theta_c$ is around $\sim 79^{\circ}$ and for the $1s$ state it is $ > 80^{\circ}$. 
Thus, the angular variation of $\tau_{2s}^{(t)}$ for $\theta < 75^{\circ}$ is exclusively due to the $\tau_{e-e}(\theta)$ contribution absent in direct ionization.
For both direct ionization as well as shake-up ionization the additivity rule [Eq.~\eqref{eq:time_delay}] applies as confirmed by comparison between the full numerical solution (symbols) and the analytic prediction (dashed lines) [Eqs.~\eqref{eq:tau_cc_pt},~\eqref{eq:dipole_delay}].
Note that we use the asymptotic prediction to calculate $\tau_{\cc}$ entering the additivity rule since for ionization from $s$ states the $\theta$ dependence of the continuum-continuum delay is negligible for $\theta < 75^{\circ}$ in the investigated energy range.
\newline
The $n=2$ correlation satellite comprises 4 degenerate ionic final states $2s,\: 2p_0, \: 2p_{\pm 1}$.
Due to electron correlation the final residual ion can not only be in the He$^+\left(2p_0 \right)$ state, but also in the He$^+\left(2p_{\pm 1} \right)$ state, as only the total magnetic moment of the atom, $M=m_1+m_2$, but not the individual magnetic momenta of the electrons, ($m_1,m_2$), is conserved for linearly polarized XUV and IR fields.
Thus, the RABBITT traces for He$(2p$) contain an incoherent sum over these substates (Fig.~\ref{fig:he_n2_shakeups}).
We observe that $\tau^{(t)}$ for the full $2p$ shell coincides with the time delay of the $2p_0$ state for low emission angles where the $2p_{\pm1}$ states have a nodal line at $\theta=0^{\circ}$.
For larger emission angles, however, the latter states become more important and the averaged time delay increasingly differs from that of the He$^+\left(2p_{0}\right)$ state. \\
We find that the analytic prediction for $\tau^{(t)}(\theta)$ [Eq.~\eqref{eq:time_delay}] (dashed lines in Fig.~\ref{fig:he_n2_shakeups}) coincides quite well with the result obtained from the simulation for all $n=2$ final states.
For small emission angles the agreement is almost perfect.
Separating the analytic prediction for the full $n=2$ shell into its different components, we find that for angles below $\theta_c$ the by far dominant contribution to the angular variation is given by $\tau_{e-e}(\theta)$.
For the cc phase we use the analytic approximation for  $\tau_{\cc}^{\mathrm{asym}}$ [Eq.~\eqref{eq:tau_cc_pt}] \cite{DahLhuMaq2012}.
We speculate that the residual differences between the analytic prediction and the full numerical result is due to the asymptotic approximation to $\tau_{\cc}$ which neglects the residual angle dependence of continuum-continuum transitions in the Coulomb field.
The simulations lie systematically  below the analytic prediction for $\theta \geq 20^{\circ}$, in line with our observation of the angle dependence of $\tau_{\cc}$  for ionization of a $p$-shell electron and recent results presented in literature \cite{HeuJimCir2016,IvaKhe2017,BusVinZho2018}.
The overall good agreement between the simulation and the analytic prediction, nevertheless, offers several qualitative insights.
First, the photoionization delay obtained by RABBITT for polarizable targets can, similarly to attosecond streaking, be separated into three different contributions.
Second, the two-electron delay $\tau_{e-e}$ is the by far dominant contribution to the angle dependence of $\tau^{(t)}$ for angles $\theta < 60^{\circ}$ where the cross section is still sizeable.
Third, the angle dependence of $\tau_{e-e}$ is well captured by Eq.~\eqref{eq:dipole_delay}.
\begin{figure}[tb]
\centering
\includegraphics[width=1\columnwidth]{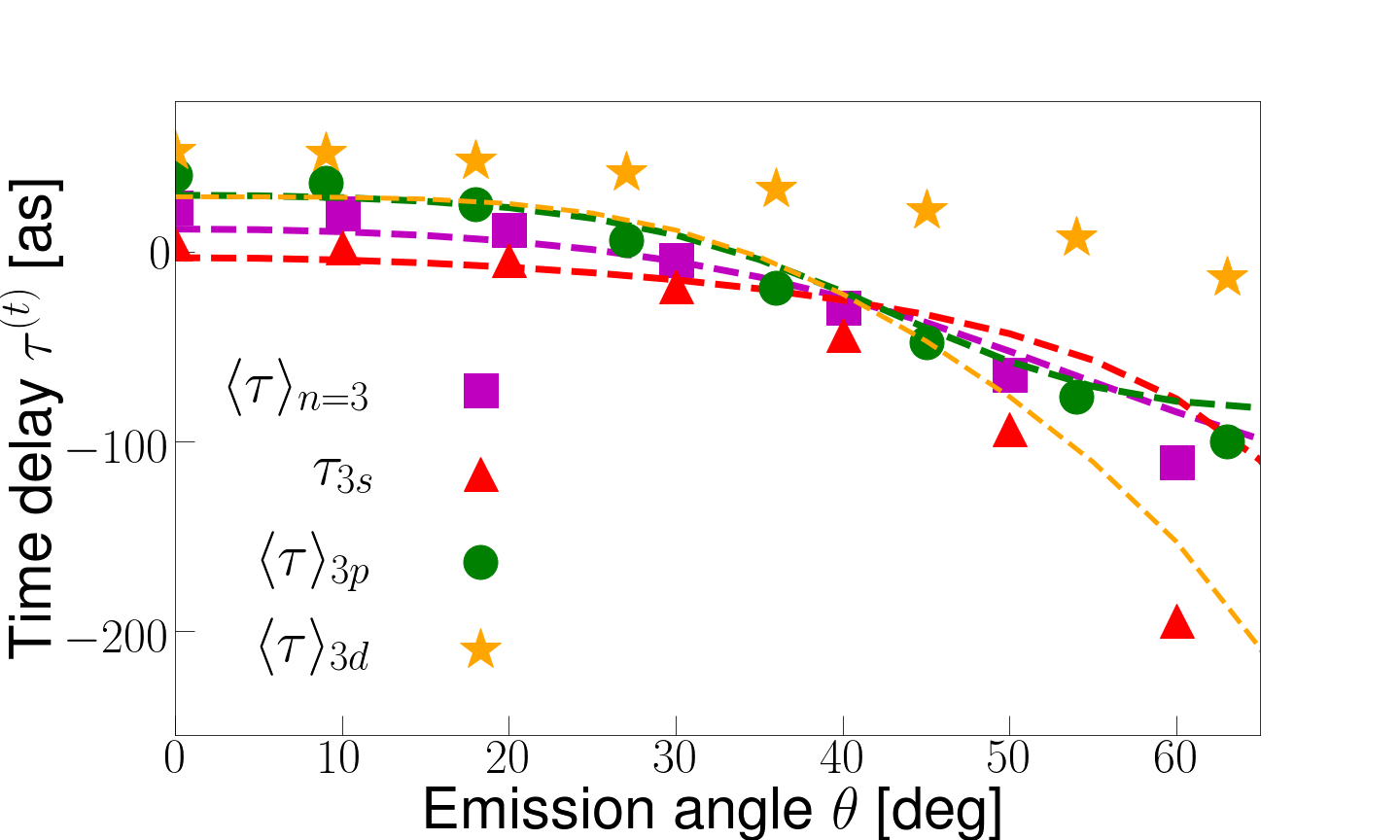}
\caption{
Time delays for the $n=3$ shake-up channels for the lowest energetic sideband (E=20.9~eV).
$\left< \tau \right>_{n=3}\left(  \theta \right)$ for the full $n=3$ shell, and resolved for the angular momentum subshells $\left< \tau \right>_{3p}$ and $\left< \tau \right>_{3d}$.
Dashed lines represent the result from Eq.~\eqref{eq:time_delay}.
} \label{fig:he_n3_shakeups}
\end{figure}
\newline
The same qualitative trends can be observed for the $n=3$ shake-up channels (Fig.~\ref{fig:he_n3_shakeups}).
Similar to the He$^{+}(n=2)$ shell, the retrieved time delays for the $n=3$ shake-up channels decrease monotonically with increasing emission angle $\theta$.
This qualitative trend is also reproduced by the analytic prediction. 
With increasing $n$ the magnitude of the delay substantially increases.
For $n=3$ values of the order of $200$~as are reached at intermediate angles well below $\theta_c$.
Unlike for $n=2$, we observe systematic deviations between the approximate analytic predictions and the numerical results.
Most notable are the differences for the $3d$ shake-up channel already in forward direction $(\theta=0^{\circ})$.
One possible explanation is the energetic proximity of the He$^{+}(n=3)$ shake-up channels to the He$^{+}(n=4)$ channels ($\Delta E \approx 2.6$eV).
Such a near degeneracy can introduce an additional intershell dipole coupling contribution to the retrieved time delays \cite{Paz2013,FeiZatNag2014}.
The analytic prediction for $\tau_{e-e}$ [Eq.~\eqref{eq:dipole_delay}] includes, however, only the intrashell coupling to the IR field.
Nevertheless, the qualitative trend of the angle dependence of $\tau^{(t)}$ is well captured by the analytic prediction for the averaged $n=3$ shell and the $3p$ shell, as well as for the $3s$ shell for $\theta<40^{\circ}$.
Again, $\tau_{e-e}$ is found to be the by far dominant contribution to $\tau^{(t)}$.

\section{From angular resolved to angle integrated time delays}\label{sec:integrated_vs_resolved}
\begin{figure}[tb]
	\begin{center}
	\includegraphics[width=1\columnwidth]{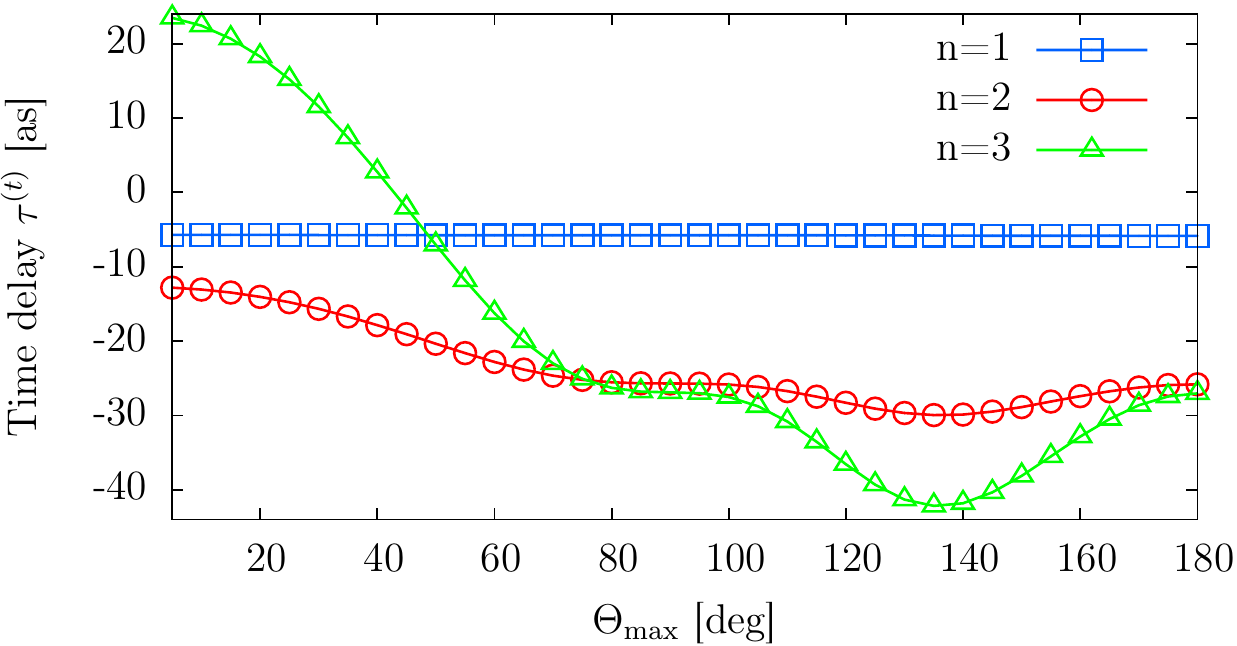}
	\end{center}
	\caption{ Total time delays as a function of the maximum collection angle $\Theta_{\mathrm{max}}$  for direct $(n=1)$ ionization of helium and ionization accompanied by shake-up to $n=2$ and $n=3$.
		The final electron energy is $\sim$~69.2 eV for the He$^+(n=1)$ channel, $\sim$~28.4~eV for the He$^+(n=2)$ channel and $\sim$~20.9~eV for the He$^+(n=3)$ channel.} \label{fig:delay_integrated}
\end{figure}
\begin{figure}[tb]
\includegraphics[width=1\columnwidth]{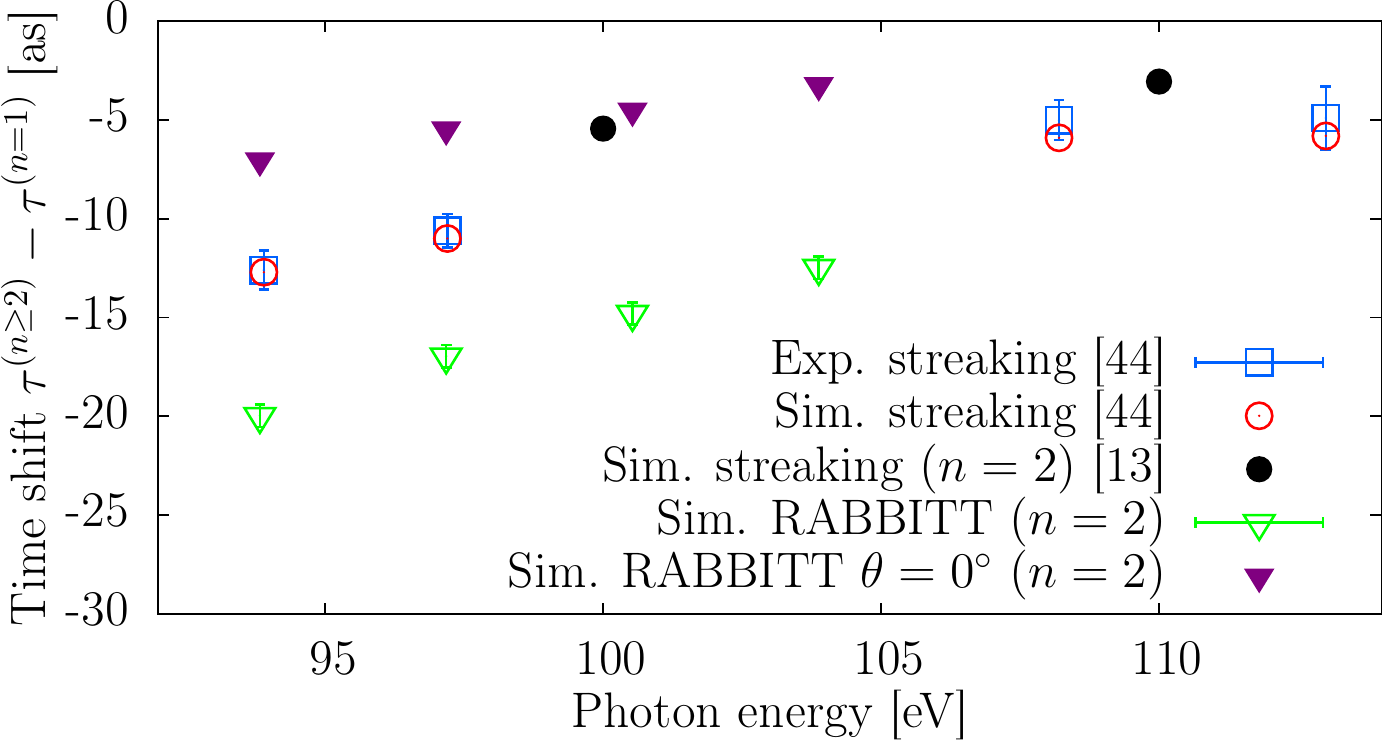}
\caption{
 Relative photoionization time delay  $\tau^{(n\geq2)}-\tau^{(n=1)}$ between electrons in the shake-up channels [He$^{+}\left(n>2\right)$] and direct ionization [He$^{+}\left(n=1\right)$] for streaking and RABBITT.
 In RABBITT the shake-up channel He$^{+}\left(n=2\right)$ is spectrally well isolated. Angle-integrated RABBITT traces (open green triangles), simulations for $n=2$ streaking (black dots), RABBITT traces evaluated in forward direction, i.e. $\theta=0^{\circ}$ (filled purple triangles).
} \label{fig:delays_rabbitt_streaking_exp}
\end{figure}

Until recently, RABBITT experiments were mostly conducted by collecting all emitted photoelectrons with a magnetic-bottle spectrometer and, thus, integrating over all emission angles.
To connect these results to the angular resolved time delays investigated in this work, we present now photoionization time delays for RABBITT traces partially integrated over angles up to an opening angle $\Theta_{\mathrm{max}}$ for atomic helium. 
We note that in many streaking experiments electrons are collected in a cone with opening angle $\Theta_{\mathrm{streak}}$ around the IR polarization direction \cite{OssSieShi2017,BisShuFor2020}, which resembles RABBITT spectra integrated up to that angle.
\\
Time delays for direct ionization show no angle dependence for $\theta<80^{\circ}$.
The steep phase drop for $\theta > 80^{\circ}$, however, is associated with a vanishingly small cross section. 
Therefore the $n=1$ time delays for partially integrated RABBITT traces are constant as a function of  $\Theta_{\mathrm{max}}$  (Fig.~\ref{fig:delay_integrated}).
A drastically different picture emerges for the correlation satellites.
The partially angle integrated spectra show a pronounced dependence on the maximum angle of integration $\Theta_{\mathrm{max}}$.
While $\tau^{(t)}$ decreases monotonically up to $\Theta_{\mathrm{max}} \simeq 140^{\circ}$ for $n=2$ and $3$, its value increases for larger $\Theta_{\mathrm{max}}$ approaching at $\Theta_{\mathrm{max}}=180^{\circ}$ the value at  $\Theta_{\mathrm{max}}=90^{\circ}$.
The reason for this is that for sufficiently long APTs and IR pulses RABBITT traces are forward-backward symmetric with respect to the electron emission angle (i.e. for $\theta \rightarrow \pi - \theta$), due to the interference between partial waves with the same parity in the sidebands.
If very short pulses or APTs consisting of even and odd harmonics were used \cite{LauCaoLi2012} this symmetry would be broken. \\
Partially integrated time delays from a RABBITT protocol also allow a direct comparison with time delays extracted from a streaking protocol. 
In general these two protocols give access to different observables.
Attosecond streaking is strongly directional collecting only electrons within an emission cone with typical opening angle, $\Theta_{\mathrm{streak}}<15^{\circ}$, about the forward direction colinear with the polarization axis.
For delays that are $\Theta_{\mathrm{max}}$ independent such as for the direct ionization of helium (Fig.~\ref{fig:delay_integrated}) streaking and angle-integrated RABBITT yield the same result for $\tau^{(t)}$.
For correlation satellites significant differences are expected.
The difference between the time delay for shake-up ($n \geq 2$) and direct ionization ($n=1$), $\tau^{(n\geq2)}-\tau^{(n=1)}$, considerably varies with the extraction protocol utilized (Fig.~\ref{fig:delays_rabbitt_streaking_exp}).
The angle-integrated and angle-resolved ($\theta=0^{\circ}$) values differ by $\simeq$~13~as at a photon energy of 90~eV.
We also find a striking difference of approximately $7$~as between the angle-integrated RABBITT data and the streaking results (both, theory and experiment)  \cite{OssSieShi2017}.
This difference is not caused by the slightly different IR wavelength used by Ossiander et al. \cite{OssSieShi2017} (800~nm), which alters the time delay caused by the IR transition ($\tau_{\cc}$ or $\tau_{\mathrm{CLC}}$) by less than 1 as.
Rather the difference can be attributed to two distinct effects.
First, due to the spectral width of the ionizing XUV pulse in streaking, it is experimentally not possible to completely separate the contributions from the  He$^{+}\left(n=2\right)$ and He$^{+}\left(n>2\right)$ channels, see Fig.~\ref{fig:shake_up_trace_740nm}a.
This admixture lowers the effective time shift $\tau^{(n\geq2)}-\tau^{(n=1)}$ compared to $\tau^{(n=2)}-\tau^{(n=1)}$ \cite{PazFeiNag2012}.
Second, the retrieved time delay is strongly $\theta$ dependent for the shake-up channels (see also Fig.~\ref{fig:delay_integrated}) and, thus, a difference between the time delay obtained from an angle-integrated RABBITT trace  and a streaking trace evaluated in forward direction ($\theta=0^{\circ}$) has to be expected.
Consequently, only the time delay difference $\tau^{(n=2)}_R-\tau^{(n=1)}_R$ obtained from RABBITT traces evaluated in forward direction ($\theta=0^{\circ}$) agrees with streaking calculations for the same quantity, i.e. for the isolated $n=2$ shake-up channel \cite{PazFeiNag2012}.
Even though completely disentangling the different shake-up channels is possible for streaking only in the simulation, the agreement between the independent RABBITT and streaking calculations confirms that the two methods do, indeed, accurately measure the same quantity for a complex multi-electron system.

\section{Concluding remarks}\label{sec:conclusion}

We have shown that the RABBITT protocol is well suited to analyze the time delay in photoionization as a  function of the emission angle of the ejected electron relative to the polarization direction for shake-up ionization of helium.
The exquisite spectral resolution allows spectral selection of different correlation satellites with residual ionic states He$^+(n)$ for $n=2,3$. 
We find that the angular variation of the delay is much more pronounced for shake-up channels than for the main line of direct ionization.
This is due to the two-electron contribution to the time delay, $\tau_{e-e}$, by which the dipolar interaction of the shaken-up polarizable bound electron with the IR field imprints an additional phase on the two-electron wave function which manifests itself as an additional phaseshift of the ionized electron.
The variation of $\tau_{e-e}$ with $\theta$ dominates over the angle variation of the EWS or cc delays at angles where the cross section is still sizeable.
This contribution therefore leaves its mark on the angle-integrated time delay.
Our numerical simulations confirm that the additivity rules for different time delays extends to the two-electron contribution.
Our present results are expected to be applicable to more complex multi-electron systems, in particular molecules featuring permanent dipole moments \cite{BagMad2010,BisShuFor2020}.

\begin{acknowledgments}
We wish to acknowledge helpful discussions with Fabian Lackner, Luca Argenti, Jan Marcus Dahlstr\"om, and David Busto.
We gratefully acknowledge funding by the FWF-DK 1243, the WWTF via project MA14-002, and the COST Action CA18222 - Attosecond Chemistry.
S. Donsa thanks the International Max Planck Research School of Advanced Photon Science for financial support.
The calculations have been performed on the Vienna Scientific Cluster (VSC).

\end{acknowledgments}

\appendix

\section{Numerical details}

In this appendix we briefly provide information on the numerical details of the calculations shown in the main text.
For more details see \cite{Don2019}.
\subsection{Single-active electron calculations}

We employ a pseudo-spectral method  \cite{NagPazFei2011,TonChu1997} to solve the time-dependent Schr\"odinger equation (TDSE) in length gauge, and expand the three-dimensional wave function into spherical harmonics.
The maximal size of the radial box was 4417 a.u., where the radial degree of freedom is discretized using the finite-element discrete variable representation (FEDVR) using up to 15 finite elements for each FEDVR element spanning 4 a.u. close to the core and 5 a.u. for $r>24$ a.u.
An absorbing boundary was used to avoid reflections of the wave function at the boundary.
We achieved converged results when including angular momenta up to $L_{\mathrm{max}}=8$.\\
The short-ranged Yukawa potential is given by 
\begin{equation}
V_{\mathrm{Y}}(r)=-\frac{1.90831}{r}e^{-r},
\end{equation}
with an ionization potential of 0.5 a.u.
For the helium and neon single-active electron calculations we use a pseudo-potential which correctly describes the ionization potential \cite{TonLin2005}.
To analyze the angle dependence of the continuum-continuum delay $\tau_{\cc}$ for the ionization of a $d$ shell electron we design a single-active electron potential, where the initial $3d$ state is energetically well separated from all other bound states.
The latter potential is given by
\begin{equation}\label{app_eq:model_pot}
    V(r)=-\frac{1+10e^{-16r}+14e^{-1.2r}}{r}.
\end{equation}
The energetically lowest bound states of this potential are given in Tab.~\ref{app_tab:states_model}. \\
The full-width at half maximum duration of the IR (XUV) pulses was chosen to be 20~(15)~fs and the corresponding peak intensities were well in the perturbative regime, i.e. $I_{\IR}=2\times 10^9$~W/cm$^2$ and $I_{\XUV}<2\times10^{11}$~W/cm$^2$.

\begin{table}[h]
\begin{center}
\begin{tabular}{c||c|c|c|c}
n & $s$ & $p$  & $d$ & $f$    \\
\hline
\hline
 1 & -282.45  & -  & - & - \\
 2 & -80.03  & -81.36  & - & - \\
 3 & -16.82  & -16.73  & -24.89 & - \\
 4 & -3.02  & -2.57  & -2.02 & -0.86\\
 5 & -1.36  & -1.21  & -1.02 & -0.56\\
\end{tabular}
\caption{\label{app_tab:states_model} Energy in eV of the lowest energetic eigenstates of the model potential given by Eq.~\eqref{app_eq:model_pot}.}
\end{center}
\end{table}

\subsection{Parameters for the \textit{ab initio} helium calculations}

For the helium shake-up calculations we use two-electron calculations from first principles \cite{FeiNagPaz2008,DonBreNi2019}.
Briefly, we solve the six-dimensional TDSE for atomic helium using the time-dependent close-coupling expansion and discretizing the radial wave functions on a spatial FEDVR grid. 
For the temporal propagation we employ the short-iterative Lanczos algorithm with adaptive time-step control. 
Spectral information is extracted by projecting the six-dimensional wave function onto products of uncorrelated Coulomb wave functions.
We use an asymmetric box where the bigger (smaller) radial size is 3857 a.u. (37 a.u.) with 11 basis function for every finite element spanning 4 a.u. close to the core and 5 a.u. for $r>$ 24 a.u.
To avoid reflection at the boundary an absorbing potential is used which starts at 3703 a.u. and 20 a.u., respectively.
Employing velocity gauge we achieve converged results for a close-coupling expansion of $L_{\mathrm{max}}=3$, $\ell_1=\ell_2=9$.  \\
We choose an IR pulse with wavelength $\lambda_{\mathrm{IR}}=$~740~nm, FWHM duration of 20~fs and peak intensity of $2\times 10^9$~W/cm$^2$.
The APT consists of the 55$^{\mathrm{th}}$, 57$^{\mathrm{th}}$, 59$^{\mathrm{th}}$, 61$^{\mathrm{st}}$, and 63$^{\mathrm{rd}}$ harmonic of the fundamental with a FWHM duration of 15~fs and peak intensities between $10^{10}$~W/cm$^2$ and $10^{12}$~ W/cm$^2$.

%


\end{document}